\documentclass{article}
\usepackage{spconf,amsmath,graphicx}

\usepackage{cite}
\usepackage{amsmath,amssymb,amsfonts}
\usepackage{algorithmic}
\usepackage{graphicx}
\usepackage{textcomp}
\usepackage{xcolor}
\usepackage{makecell}
\usepackage{booktabs}
\usepackage{tikz}
\usepackage{pgfplots}
\usepackage{hyperref}

\def\BibTeX{{\rm B\kern-.05em{\sc i\kern-.025em b}\kern-.08em
		T\kern-.1667em\lower.7ex\hbox{E}\kern-.125emX}}
\usepackage{bm}		

\def\inputImage{\ensuremath{\bm{x}}}
\def\outputImage{\ensuremath{\bm{\hat{\inputImage}}}}
\def\latentRepresentation{\ensuremath{\bm{y}}}
\def\intermediateLatentRepresentation{\ensuremath{\bm{u}}}
\def\intermediateLatentRepresentationDec{\ensuremath{\bm{v}}}
\def\encoderNet{\ensuremath{f_\mathrm{enc}}}
\def\decoderNet{\ensuremath{f_\mathrm{dec}}}
\def\codec{\ensuremath{f}}
\def\mask{\ensuremath{\bm{m}}}
\def\loss{\ensuremath{\mathcal{L}}}
\def\lossHVS{\ensuremath{\loss_\mathrm{HVS}}}
\def\lossTask{\ensuremath{\loss_\mathrm{VCM}}}
\def\rate{\ensuremath{R}}
\def\distortion{\ensuremath{D}}
\def\lagrange{\ensuremath{\lambda}}
\def\weights{\ensuremath{\bm{\theta}}}
\def\width{\ensuremath{W}}
\def\height{\ensuremath{H}}

\newcommand\copyrighttext{%
	\footnotesize Accepted for publication in 2023 International Conference on Acoustics, Speech, and Signal Processing (ICASSP). Personal use of this material is permitted.
	Permission from IEEE must be obtained for all other uses, in any current or future 
	media, including reprinting/republishing this material for advertising or promotional 
	purposes, creating new collective works, for resale or redistribution to servers or 
	lists, or reuse of any copyrighted component of this work in other works. 
}

\newcommand\copyrightnoticeOwn{%
	\begin{tikzpicture}[remember picture,overlay]
		\node[anchor=north,yshift=-10pt] at (current page.north) {\fbox{\parbox{\dimexpr\textwidth-\fboxsep-\fboxrule\relax}{\copyrighttext}}};
	\end{tikzpicture}%
	\vspace{-8mm}
}

\ninept

\newcommand{\showmark}[2]{
	\begin{tikzpicture}[baseline, baseline=-.5ex]	
		\draw[#1](0,0) -- (5mm,0); 
		\node[#1, mark size=3, mark options={solid}] at (2.5mm,0){%
			\pgfuseplotmark{#2}%
		};
	\end{tikzpicture}%
}

\newcommand{\showmarkDashed}[2]{
	\begin{tikzpicture}[baseline, baseline=-.5ex]	
		\draw[#1, dashed](0,0) -- (5mm,0); 
		\node[#1, mark size=3, mark options={solid}] at (2.5mm,0){%
			\pgfuseplotmark{#2}%
		};
	\end{tikzpicture}%
}

\definecolor{color0}{rgb}{0.12156862745098,0.466666666666667,0.705882352941177}
\definecolor{color1}{rgb}{1,0.498039215686275,0.0549019607843137}
\definecolor{color2}{rgb}{0.172549019607843,0.627450980392157,0.172549019607843}
\definecolor{color3}{rgb}{0.83921568627451,0.152941176470588,0.156862745098039}
\definecolor{color4}{rgb}{0.580392156862745,0.403921568627451,0.741176470588235}
\definecolor{color5}{rgb}{0.549019607843137,0.337254901960784,0.294117647058824}
\definecolor{color6}{rgb}{0.890196078431372,0.466666666666667,0.76078431372549}

\definecolor{blueLight}{HTML}{8c564b}
\definecolor{blueMedium}{HTML}{bcbd22}
\definecolor{blueDark}{HTML}{17becf}

\title{Saliency-Driven Hierarchical Learned Image Coding for Machines}
\name{Kristian Fischer, Fabian Brand, Christian Blum, and Andr\'e Kaup\thanks{The authors gratefully acknowledge that this work has been funded by the Deutsche Forschungsgemeinschaft (DFG, German Research Foundation) under project number 426084215.}}
\address{Multimedia Communications and Signal Processing\\Friedrich-Alexander-Universit\"at Erlangen-N\"urnberg (FAU)}

\begin{document}
	%
	\maketitle
	\copyrightnoticeOwn
	\begin{abstract}
		We propose to employ a saliency-driven hierarchical neural image compression network for a machine-to-machine communication scenario following the compress-then-analyze paradigm.
		By that, different areas of the image are coded at different qualities depending on whether salient objects are located in the corresponding area.
		Areas without saliency are transmitted in latent spaces of lower spatial resolution in order to reduce the bitrate.
		The saliency information is explicitly derived from the detections of an object detection network.
		Furthermore, we propose to add saliency information to the training process in order to further specialize the different latent spaces.
		All in all, our hierarchical model with all proposed optimizations achieves 77.1\,\% bitrate savings over the latest video coding standard VVC on the Cityscapes dataset and with Mask R-CNN as analysis network at the decoder side. 
		Thereby, it also outperforms traditional, non-hierarchical compression networks.
	\end{abstract}

	\section{Introduction}
	
	In today's modern world of communication, the number of applications and processes where machines and devices are communicating with each other has tremendously increased.
	Such communication is often referred to as machine-to-machine~(M2M) communication, taking mostly place in Internet of things~(IoT) scenarios.
	From this raise, it can be followed that suitable image and video compression schemes are necessary. 
	This has been targeted by MPEG since 2019~\cite{zhang2019}, referring to this special type of video or image coding as \textit{video coding for machines~(VCM)}.
	
	This work focuses on improving the image coding performance for instance segmentation networks as information sink following the compress-then-analyze paradigm~\cite{redondi2016}.
	For such scenarios, previous approaches~\cite{galteri2018, choi2018, fischer2021_ICASSP} for standard hybrid codecs mainly utilized saliency coding by separating the image into salient and non-salient areas.
	The latter ones are coded with reduced bitrate without harming the accuracy of the analysis network. 
	More recent work~\cite{chamain2021,le2021_ICASSP, fischer2022_journal} employed neural image compression networks~(NCNs) as codecs, which allows for an end-to-end training of the whole VCM framework with the analysis network as discriminator.
	By that, the NCNs are adapted towards the characteristics of the analysis network and the input data which eventually results in a superior coding performance outperforming VVC~\cite{bross2021_VVC} for the tasks of object detection and segmentation as it has been shown in \cite{le2021_ICASSP} and \cite{fischer2022_journal}.
	
	\begin{figure}[t]
		\centering
		\includegraphics[width=0.83\linewidth]{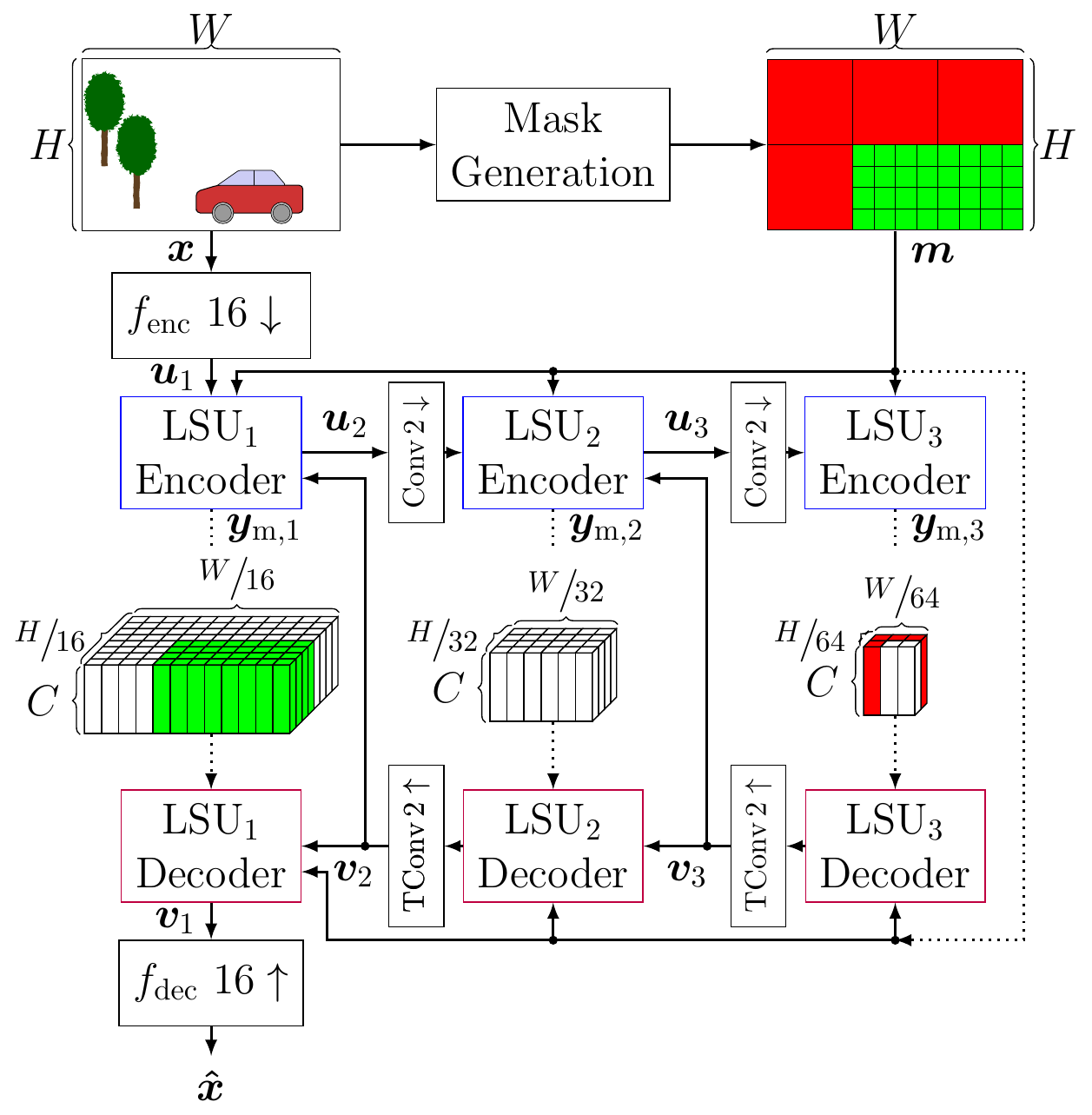}
		\vspace{-2mm}
		\caption{Proposed hierarchical neural image coding framework based on the RDONet structure~\cite{brand2021_CVPR, brand2022} for VCM scenarios. 
			The input image $\inputImage$ is separated into salient (green) and non-salient (red) areas that are transmitted in different latent spaces $\latentRepresentation_n$. 
			White latent elements correspond to a value of zero since these latents are masked out.
			Dotted lines indicate that this information is transmitted via the channel.
			$\downarrow s$ and $\uparrow s$ indicate a {down-upscaling} factor of $s$. 
			$C$ denotes the number of latent space channels.
			Please note that $\inputImage$ and $\latentRepresentation_n$ are not in scale for better visualization.	
		}
		\label{fig:coding framework}
		\vspace{-5mm}
	\end{figure}
	
	However, the methodology in~\cite{chamain2021,le2021_ICASSP} has the shortcoming that the network has to implicitly derive between salient areas where a high quality is required, and non-salient areas where no potential objects are located. Due to the limited field of view of the NCNs, this decision has to be drawn on a rather small amount of pixels. 
	Alleviation to this was made in our previous work~\cite{fischer2022_journal}, where we introduced a latent space masking network to mask possibly non-salient areas in the latent space in order to reduce the required bitrate. 
	Nevertheless, also this information was only implicitly derived from the features of the analysis network.
	
	To overcome these issues, we propose to employ a hierarchical NCN, as it is depicted in Fig.~\ref{fig:coding framework}, which utilizes multiple latent spaces $\latentRepresentation_n$.
	These are used to compress different areas of the image with different quality.
	By that, the latent spaces are specialized for either transmitting salient or non-salient areas.
	In Fig.~\ref{fig:coding framework}, this means that areas including objects of interest, e.g. cars, are transmitted with higher spatial resolution and thus a higher quality, whereas the non-salient areas, e.g. trees, are transmitted with less spatial resolution requiring less bitrate.
	In our approach, the saliency information is explicitly derived from an external object detection network and used to steer the NCN.
	We propose to employ our existing rate-distortion optimization network~(RDONet)~\cite{brand2021_CVPR, brand2022} as the core network structure. 
	This NCN allows for hierarchical coding by transmitting the image data in three latent spaces of different spatial resolutions.
	To the best of our knowledge, we are the first to propose a learned compression framework exploiting VCM-based saliency information.
	
	
	All in all, our paper provides the following contributions: 
	First, we show that the hierarchical RDONet published in~\cite{brand2022} outperforms the latest standard hybrid video codec VVC and a comparable NCN architecture with only one latent space, when being trained in an end-to-end manner similar to~\cite{chamain2021,le2021_ICASSP, fischer2022_journal}. 
	Second, we further improve the coding performance of RDONet for VCM scenarios by proposing a new RDOnet masking criterion, which allows to explicitly add saliency information to the coding process during the inference. 
	Third, we show that the overall coding performance is further improved by adding saliency information to the training process to specialize the different latent spaces in coding salient or non-salient areas.
	
	\begin{figure}
		\centering
		\includegraphics[width=0.7\linewidth]{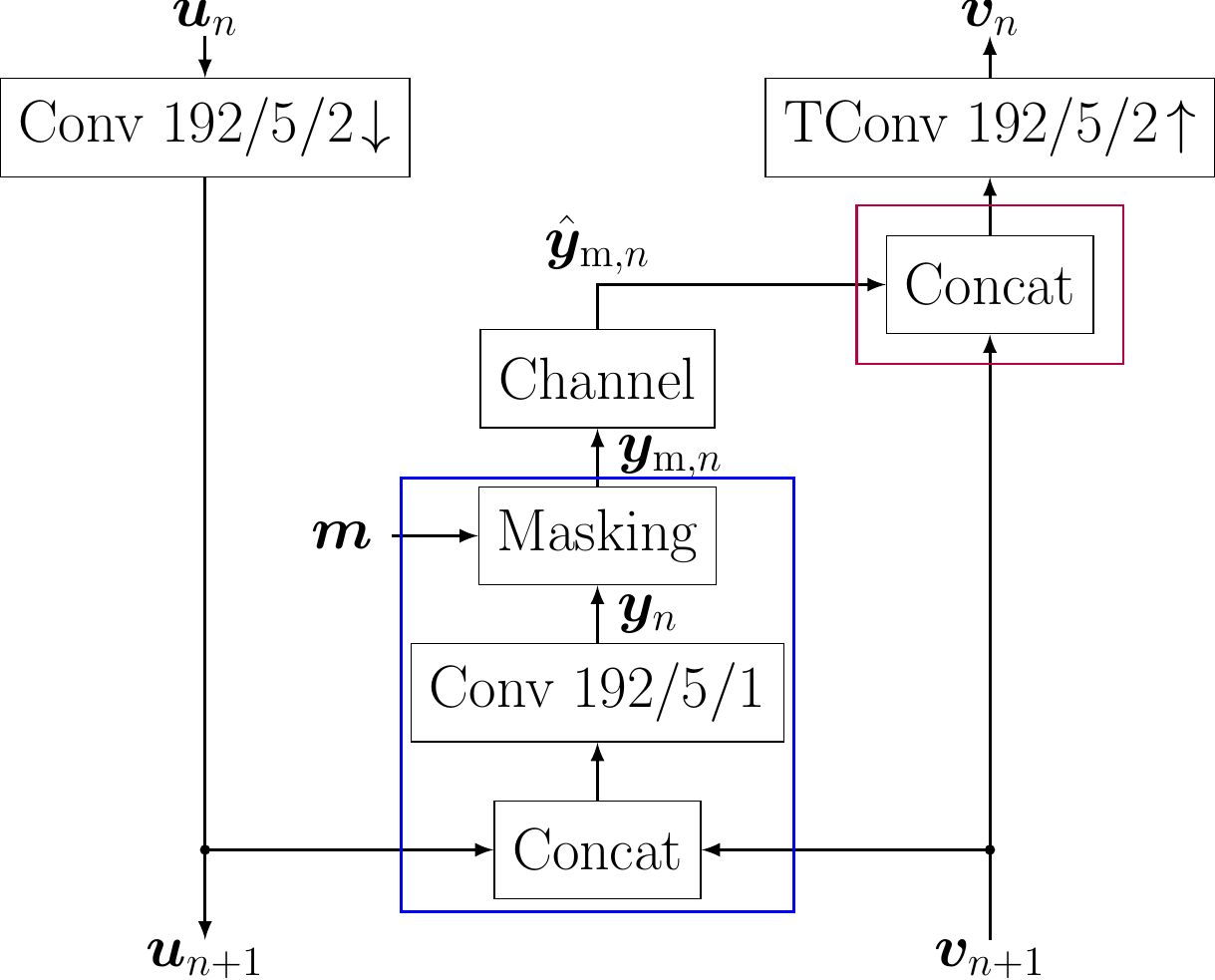}
		\vspace{-2mm}
		\caption{$n$-th LSU structure with attendant convolutional layers based on RDONet~\cite{brand2021_CVPR}. 
			The encoder and decoder part (ref. Fig~\ref{fig:coding framework}) are framed in blue and purple color, respectively.
			Conv $C/k/s$ denotes a convolutional layer with $C$ output channels, a kernel size of $k\times k$, and a subsampling factor of $s$. TConv denotes an analogous transposed convolutional layer.}
		\label{fig:lsu}
		\vspace{-4mm}
	\end{figure}
	
	\section{Hierarchical Neural Image Compression}
	Today, end-to-end trained neural image compression networks are mainly based on the pioneering work by Ball\' e et al.~\cite{balle2017endtoend}.
	There, the authors proposed a variational autoencoder \codec\ that transforms the input image \inputImage\ by an encoder network into a latent space \latentRepresentation\ of reduced spatial dimensionality that is quantized and losslessly transmitted to the decoder side. 
	The corresponding decoder network reconstructs the image from the transmitted latent space resulting in the deteriorated output image \outputImage. 
	To train the network weights \weights, a loss $\lossHVS$ combining the required rate \rate\ and the distortion \distortion\ between the input \inputImage\ and the output $\outputImage$ is utilized
	\begin{equation}
		\lossHVS = \distortion(\inputImage, \outputImage) + \lagrange \cdot \rate(\encoderNet(\inputImage| \weights)),
	\end{equation}
	where \lagrange\ steers between the two competing goals of a low bitrate and a low distortion.

	In order to provide additional transmission options, RDONet~\cite{brand2021_CVPR,brand2022} adds extra latent spaces $\latentRepresentation_n$ to the NCN structure proposed in~\cite{minnen2018_mbt2018}, which allows compressing different areas of \inputImage\ at different spatial resolutions.
	The spatial resolution is halved with every deeper latent space such that one element in the latent space covers more pixels in the image domain. Thus, more bitrate can be saved and the weights can be adapted correspondingly.
	Thereby, the external mask \mask\ steers which image area is coded by which latent space.
	Each area of \inputImage\ is transmitted by exactly one latent space.
	The non-selected areas are zeroed-out (white latents in Fig.~\ref{fig:coding framework}).
	
	The RDONet coding order is that the deepest latent space $\latentRepresentation_3$ is coded first and the coding process of each latent space $\latentRepresentation_n$ is conditioned on the before transmitted latent space $\latentRepresentation_{n+1}$.
	First, the image $\inputImage$ with size $\height\times \width$ is fed into $\encoderNet$ consisting of four convolutional layers with a stride of two in order to reduce the spatial resolution. 
	This results in the intermediate latent space $\intermediateLatentRepresentation_1$.
	The three latent spaces $\latentRepresentation_n$ are generated from feeding $\intermediateLatentRepresentation_1$ into three cascaded latent space units~(LSUs)~\cite{brand2021_CVPR}.
	The structure of the $n$-th LSU is depicted in Fig.~\ref{fig:lsu}.
	First, the incoming data $\intermediateLatentRepresentation_n$ is spatially downscaled by a convolution with a stride of two.
	The resulting features are concatenated with the output of the deeper $\mathrm{LSU}_{n+1}$ $\intermediateLatentRepresentationDec_{n+1}$ and fed into a convolutional layer to obtain the latent space $\latentRepresentation_n$.
	Subsequently, elements of $\latentRepresentation_n$ that are not transmitted in this latent space are zeroed-out depending on \mask.
	The channel to transmit the masked latents $\latentRepresentation_{\mathrm{m},n}$ is similar to~\cite{minnen2018_mbt2018}.
	$\latentRepresentation_{\mathrm{m},n}$ is quantized, coded with the help of a conditional hyperprior including a context model, and transmitted to the decoder side. 
	There, the received latent space $\hat{\latentRepresentation}_{\mathrm{m},n}$ is concatenated with the result from the deeper $\mathrm{LSU}_{n+1}$ $\intermediateLatentRepresentationDec_{n+1}$, spatially upscaled by a transposed convolution, and fed forward into the next LSU.
	After the last latent space $\latentRepresentation_1$ has been transmitted, the decoder network \decoderNet\ reconstructs the output image \outputImage\ from the output of $\mathrm{LSU}_1$ $\intermediateLatentRepresentationDec_1$.
	
	
	
	
	\section{Optimizing Hierarchical NCN Framework for VCM Scenarios}
	This section discusses the adaptations that are proposed in order to adapt RDONet for the compression of images in the VCM context. As analysis network at the decoder side, the state-of-the-art instance segmentation network Mask R-CNN~\cite{he2017} is chosen.
	
	\subsection{End-to-end Training for Analysis Network}
	In general, NCNs are optimized for the task of coding for the human visual system~(HVS).
	Originally, the weights \weights\ of RDONet are thus trained on a distortion $\distortion_\mathrm{HVS}$ mixing MS-SSIM and MSE~\cite{brand2021_CVPR}:
	\begin{equation}
		\distortion_\mathrm{HVS} = D_\mathrm{MSE} + 0.1 \cdot D_\mathrm{MS-SSIM}.
		\label{eq:loss HVS}
	\end{equation}
	
	To optimally adapt RDONet to the Mask R-CNN as information sink, we end-to-end train its weights with the analysis network as discriminator in the training loop similar to the work in \cite{chamain2021,le2021_ICASSP,fischer2022_journal}. 
	Therefore, we substitute $D_\mathrm{HVS}$ by the Mask R-CNN task loss $\loss_\mathrm{MRCNN}$~\cite{he2017} to obtain the following VCM-optimized loss:
	\begin{equation}
		\lossTask = \loss_\mathrm{MRCNN}(\outputImage) + \lagrange \cdot \rate(\codec(\inputImage| \weights)).
	\end{equation}
	The analysis network weights are not adapted during this training. 
	
	\subsection{VCM-Optimized Mask Generation for Inference}
	\label{subsec:VCM optimized mask generation}
	
\begin{figure}[!t]%
		\centering
		\begin{tabular}{p{0.45\linewidth}p{0.45\linewidth}}
			\includegraphics[width=\linewidth]{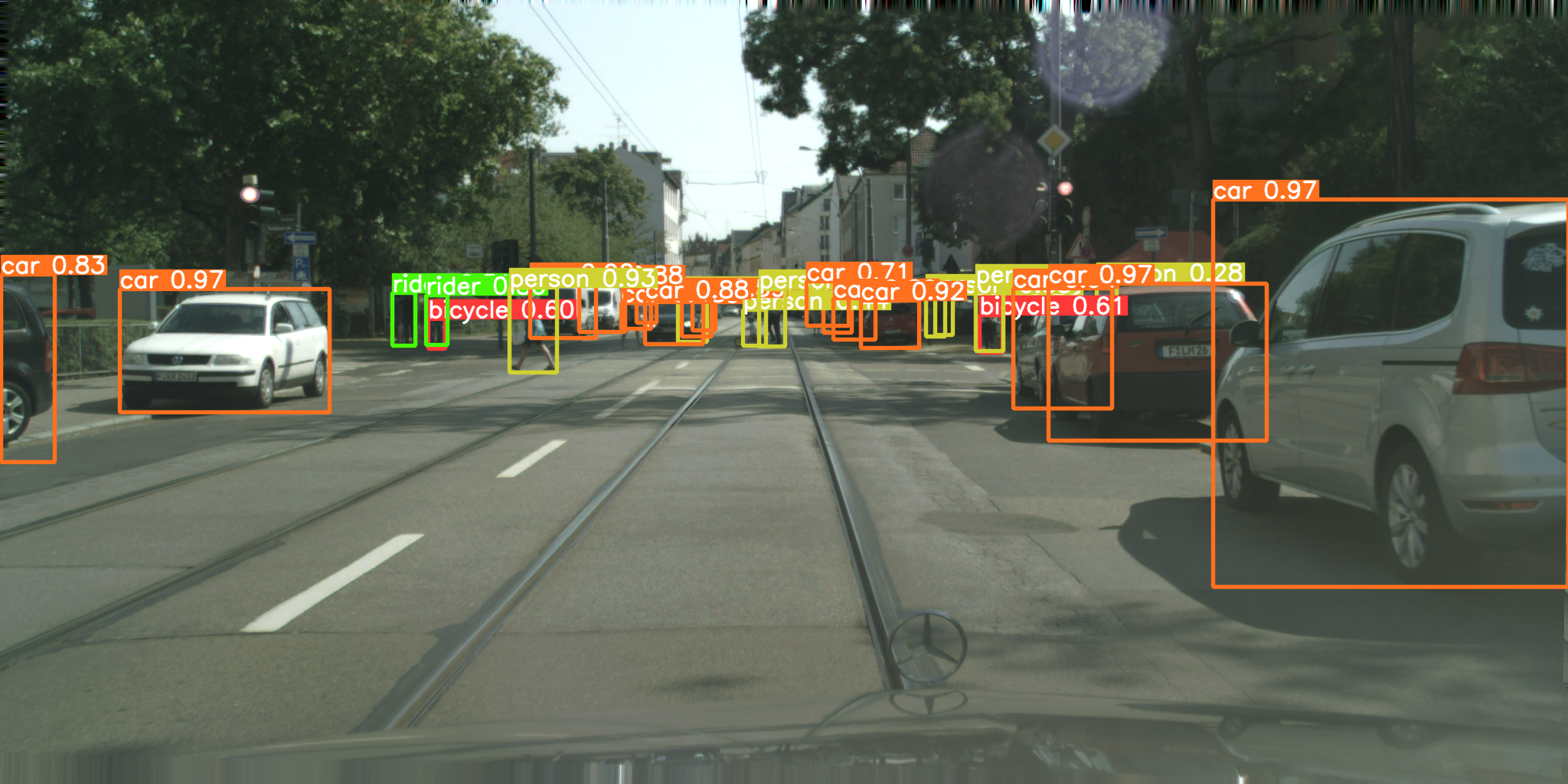} &
			\includegraphics[width=\linewidth]{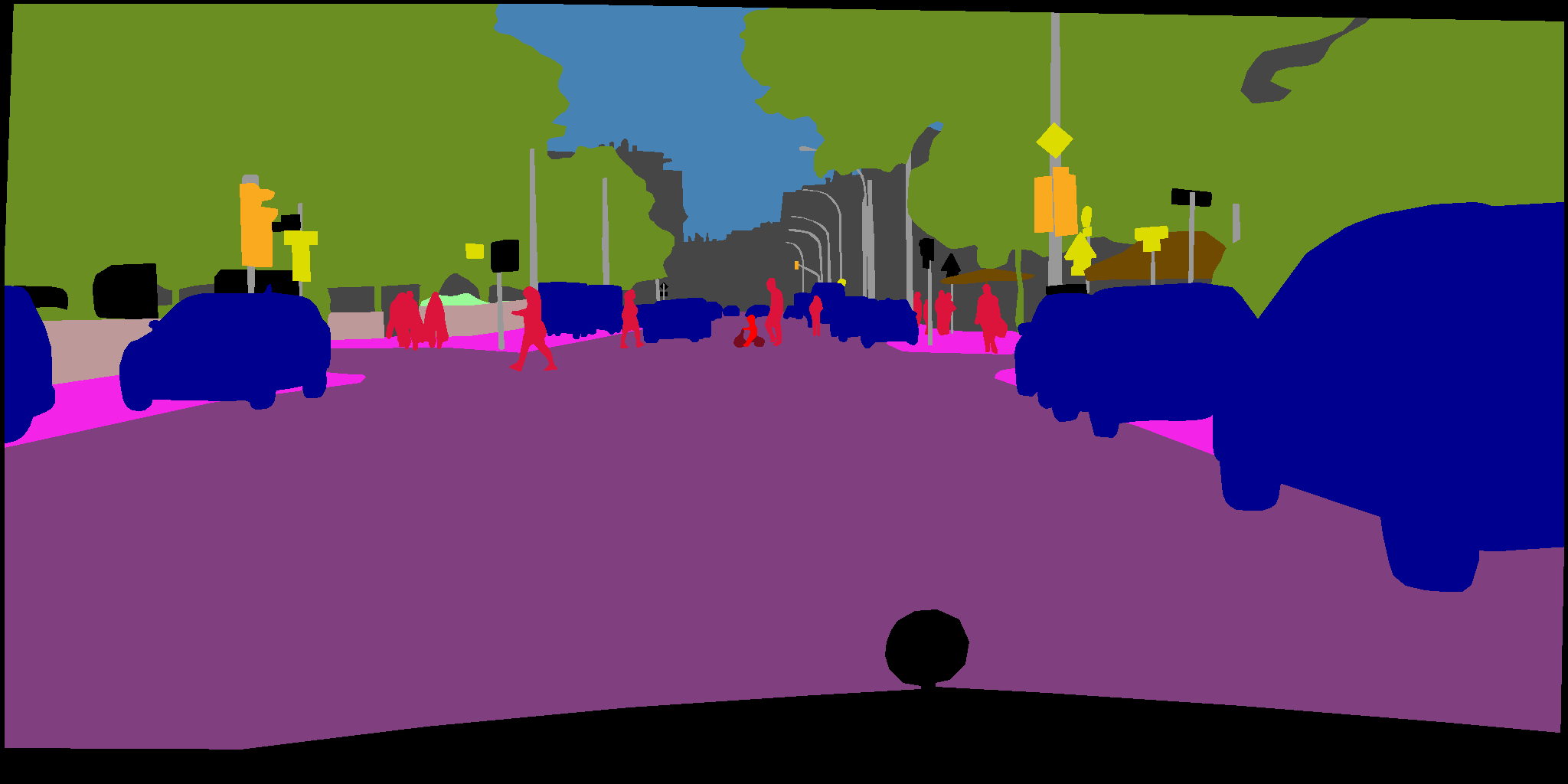} \\
			
			\footnotesize{\makecell[t]{a) Input image with \\ YOLO detections}} &
			\footnotesize{\makecell[t]{b) Annotated ground \\ truth~(GT) data}} \\
			
			\includegraphics[width=\linewidth]{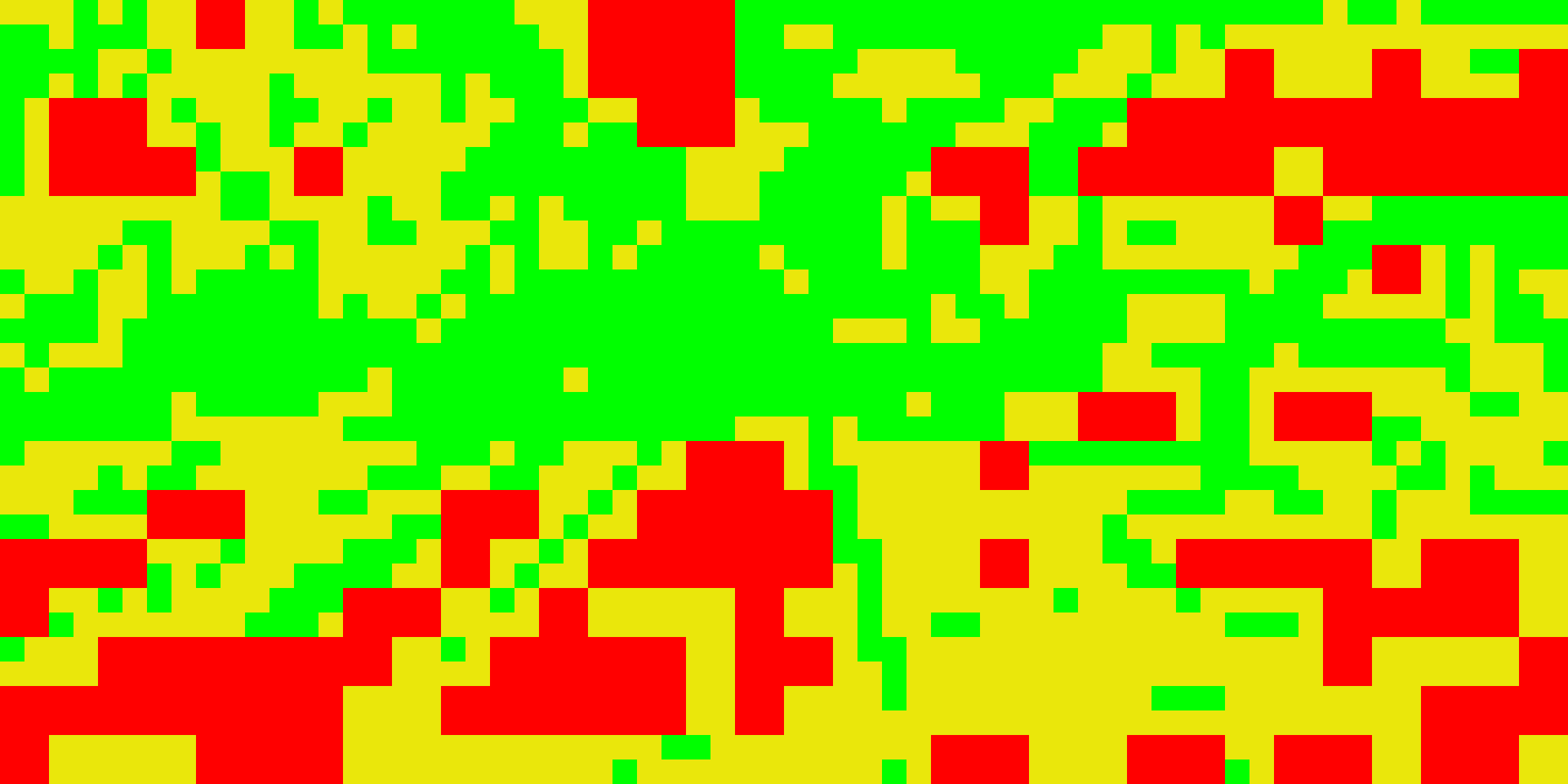} &	
			\includegraphics[width=\linewidth]{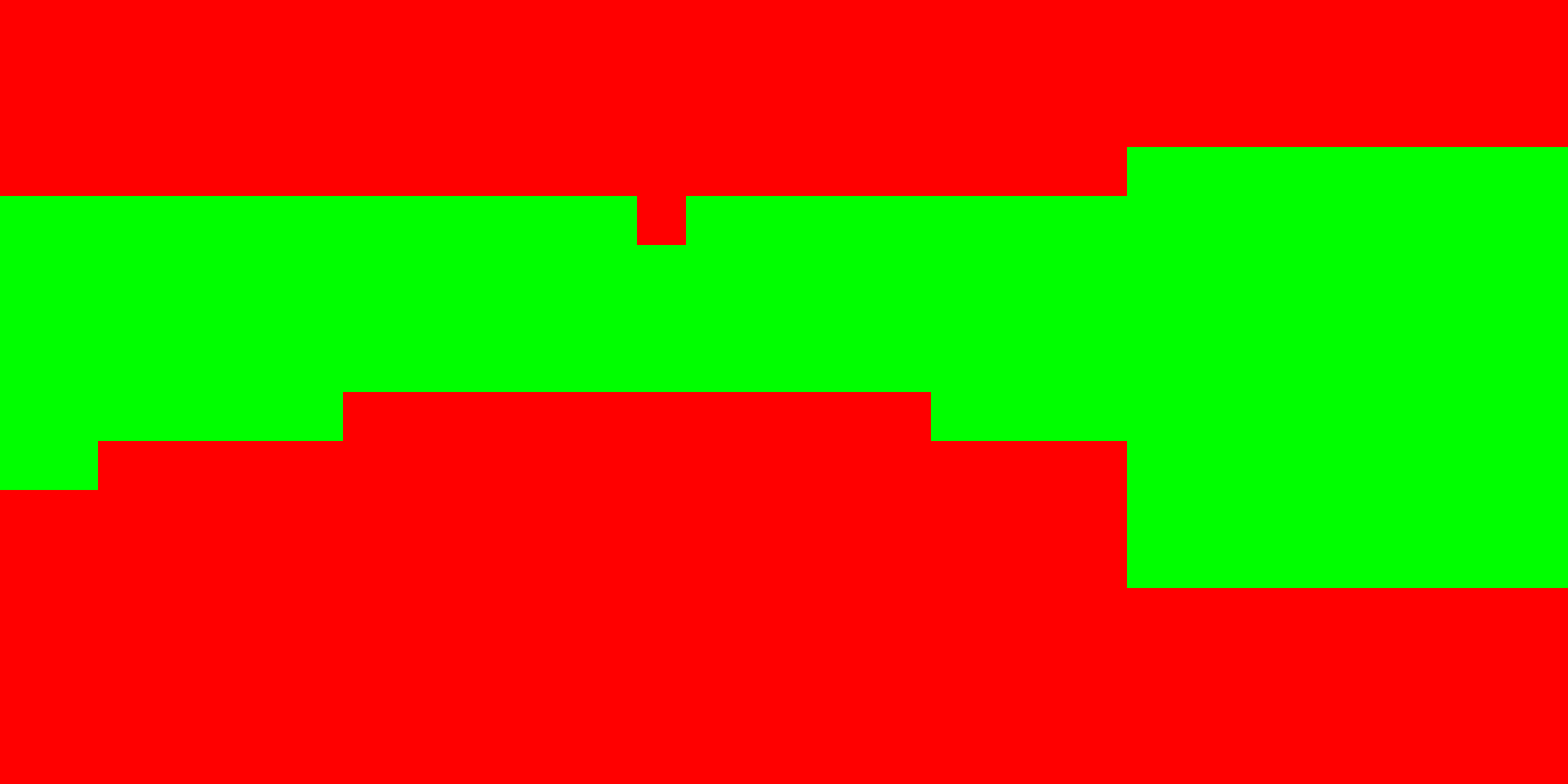} \\
			
			\footnotesize{\makecell[t]{c) Mask generated by variance \\ criterion (var mask)~\cite{brand2022}} } &
			\footnotesize{\makecell[t]{d) Mask generated by proposed \\YOLO criterion (YOLO mask)}}
		\end{tabular}
		\caption{Exemplary masks \mask\ for Cityscapes input image \textit{frankfurt\_000000\_001236\_leftImg8bit}. 
			The used color mapping is green~$\rightarrow \latentRepresentation_1$, yellow~$\rightarrow \latentRepresentation_2$, and red $\rightarrow \latentRepresentation_3$.
		}
		\vspace{-6mm}
		\label{fig:exemplary masks}%
	\end{figure}

	When applying RDOnet for the HVS,~\cite{brand2022} showed that deriving the masks based on the variance of each block is a decent compromise between rate-distortion performance and runtime.
	Fig~\ref{fig:exemplary masks}c shows such a variance-based mask.
	From this we can easily follow that the generated mask is sub-optimal, since a lot of highly structured content such as the trees or road markings would still be encoded in the first latent space $\latentRepresentation_1$ requiring a lot of bitrate, despite being not relevant for the analysis network at the decoder side.
	
	In order to obtain optimal masks for inference when coding for the task of instance segmentation, we propose to apply an object detection network to the input data \inputImage\ inspired by our previous work in~\cite{fischer2021_ICASSP}.
	There, YOLO~\cite{redmon2016_ieee} is applied to the input image to derive the salient objects, to ultimately reduce the bitrate in non-salient coding units of VVC.
	This successful criterion is transferred to our RDONet approach by transmitting all image areas that are covered by the bounding box of a YOLO detection in $\latentRepresentation_1$. 
	All remaining areas are transmitted in $\latentRepresentation_3$, since \cite{fischer2021_ICASSP} revealed that the best coding performance is achieved, when the non-salient areas are transmitted at the lowest possible quality.
	Thus, $\latentRepresentation_2$ is not utilized with this mask.
	Since the mask signaling to the decoder is very cheap in RDOnet, the bitrate overhead when keeping this second latent representation in the framework can be neglected.
	The mask generated from the YOLO detections is depicted in Fig.~\ref{fig:exemplary masks}d and shows that only the areas containing relevant objects such as cars and pedestrians are transmitted with the best quality.
	
	\subsection{VCM-Adapted Training with Ground-Truth Data}
	\label{subsec: improved training with GT data}
	With the proposed mask criterion during inference, a discrepancy arises between the mask generation in training, i.e. variance based as proposed in~\cite{brand2022}, and inference, i.e. VCM-optimized. 
	Hence, the codec cannot optimally adapt its weights to the different tasks of delivering a high quality for salient areas transmitted by $\latentRepresentation_1$, and reducing the rate in non-salient areas transmitted in $\latentRepresentation_3$. 
	
	To mitigate this discrepancy, we propose to utilize the ground truth data, which is commonly available when training with the task loss of the analysis network.
	Analogous to the masking criterion in inference with YOLO detections as presented in the previous Section~\ref{subsec:VCM optimized mask generation}, we generate a mask based on the ground truth objects.
	If a pixel of $\inputImage$ is located inside an annotated object, the corresponding block is coded in $\latentRepresentation_1$.
	All other objects are coded in latent space $\latentRepresentation_3$.
	Therewith, the network learns that all information that is transmitted in $\latentRepresentation_3$ does not influence the Mask R-CNN task loss $\loss_\mathrm{MRCNN}$, and thus reduces the bitrate in such regions as far as possible. 
	
	\vspace{-3mm}
	\section{Analytical Methods}
	\subsection{Training Procedure}
	When training our NCN models, we selected four $\lagrange$ parameters such that the coding results are in a comparable bitrate range as the reference VVC test model~(VTM-10.0)~\cite{chen2020vtm10} with the four quantization parameter~(QP) values of 22, 27, 32, and 37.
	We trained the models on the Cityscapes training dataset cropped to $512\times1024$ patches and a batch size of eight.
	As optimizer, we used Adam with a learning rate of $0.0001$.
	First, we trained a reference RDONet model on $\lossHVS$ as in \eqref{eq:loss HVS} with variance masks for 1500 epochs as described in~\cite{brand2022}.
	These weights were taken as initialization to further train the models with the proposed VCM optimizations, i.e. the training with $\lossTask$ and the training with the GT-based masks, for another 1000 epochs.
	To generate the VCM-optimized masks for inference, we trained a YOLO-v5 network~\cite{jocher2020YoloLibrary} on the Cityscapes training data for 600 epochs with the standard configuration from~\cite{jocher2020YoloLibrary}.
	
	\subsection{Evaluation Setup}
	To evaluate our proposed methods, we build up a coding framework similar to~\cite{fischer2020_ICIP} in line with the common testing conditions for VCM proposed by MPEG~\cite{liu2020_VCM_CTC}.
	As dataset, we compressed the 500 uncompressed Cityscapes~\cite{cordts2016} validation images. 
	The compressed images were taken as input for the Mask R-CNN~\cite{he2017} instance segmentation network with ResNet50 backbone.
	Its weights trained on the Cityscapes training data were taken off the shelf from the Detectron2 library~\cite{wu2019detectron2}.
	The Mask R-CNN accuracy is measured by the average precision~(AP), which is the standard metric to evaluate instance segmentation networks.
	To alleviate class imbalances, we calculate the weighted AP~(wAP) as in~\cite{fischer2020_ICIP}.
	The resulting rate-wAP curves are quantified by the Bj\o ntegaard delta rate~(BDR) metric~\cite{bjontegaard2001_new}, measuring the bitrate savings at the same detection accuracy by coding the data with a certain codec over an anchor codec.
	As a reference, we compare our methods against the VTM-10.0 and the NCN from~\cite{fischer2022_journal} with a similar codec structure but only one transmitted latent space.
	
	\vspace{-3mm}
	\section{Experimental Results}
	
	\subsection{Influence of End-to-end Training with Analysis Network}
\begin{figure}
		\centering
		\includegraphics[height=4cm]{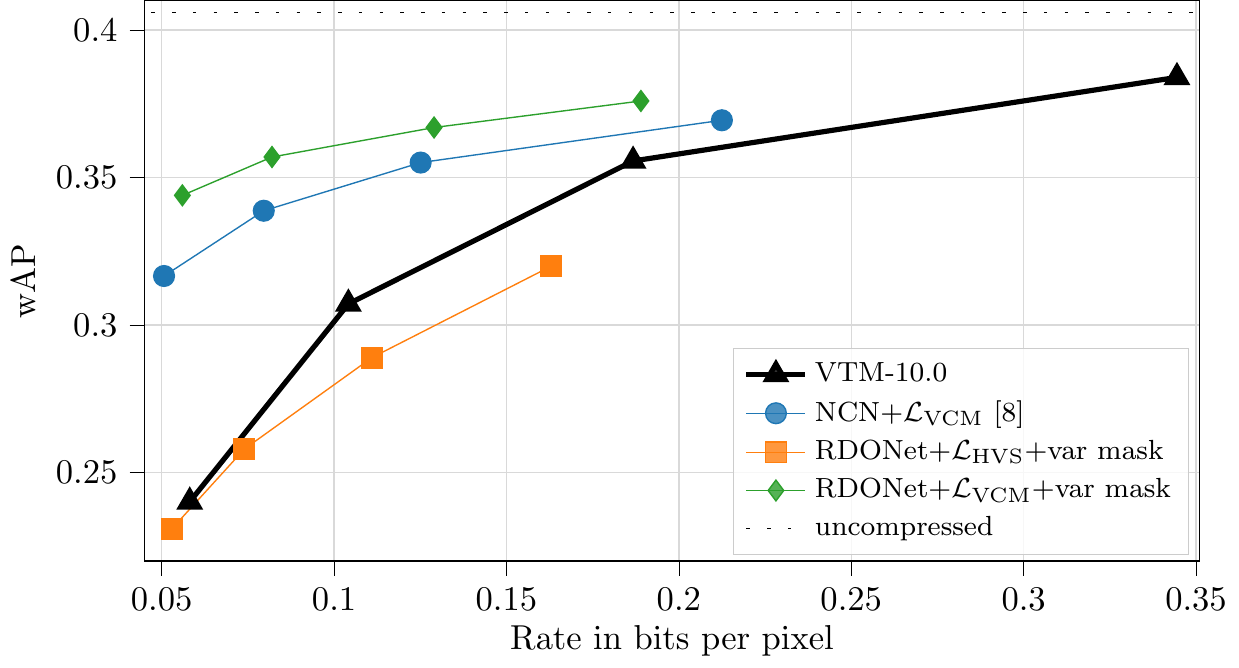}
		\vspace{-3mm}
		\caption{wAP-rate curves averaged over the 500 Cityscapes validation images. NCN denotes the reference network with only one latent space. The dotted line shows the accuracy when applying Mask R-CNN to uncompressed images.}
		\label{fig:wAP-rate curves}
		\vspace{-5mm}
	\end{figure}

	Fig.~\ref{fig:wAP-rate curves} shows the coding efficiency of the tested coding methods.
	The RDONet model trained for the human visual system~\cite{brand2022} (orange) is performing worse than VTM-10.0 in terms of wAP-rate performance.
	The reference NCN~\cite{fischer2022_journal} with one latent space trained on \lossTask\ (blue) outperforms the reference VTM-10.0 codec by 41.4\,\% BDR savings (cf. Tab.~\ref{tab:BD results}).
	The proposed approach of coding the data with the hierarchical RDONet structure steered with the masks derived from the basic variance criterion (green) results in even better coding performance of 52.7\,\%.

	\subsection{Influence of Advanced Mask Generation for Inference}
\begin{figure}[t]
		\centering
		\includegraphics[height=4cm]{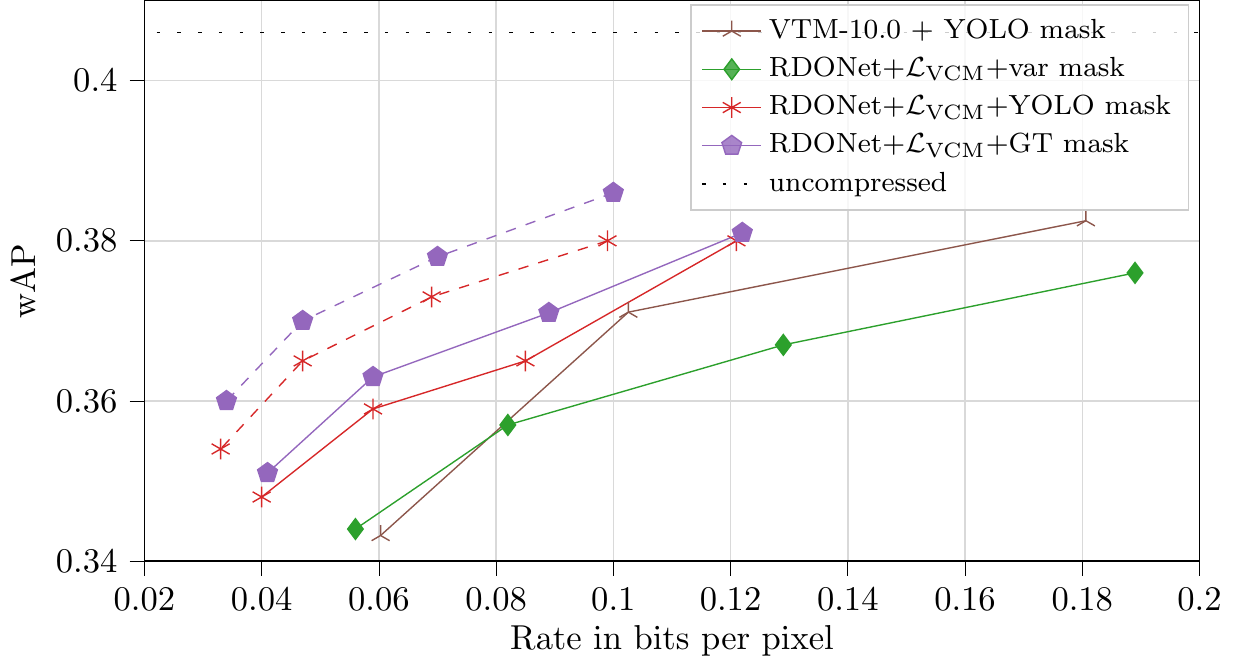}
		\vspace{-3mm}
		\caption{Comparison of the RDONet model trained with $\lossTask$ depending on the used masks during inference coding the 500 Cityscapes validation images. Solid and dashed lines symbolize that the RDONet model was trained with variance and GT-based masks, respectively.}
		\label{fig:comparison different mask generation methods}
		\vspace{-3mm}
	\end{figure}

	Next, Fig.~\ref{fig:comparison different mask generation methods} compares the coding performance employing different masking criteria to obtain $\mask$ for the inference.
	Here, the trained model remains the same and only the masks are changed during the inference.
	The reference case (green) is the mask generated by the variance criterion from~\cite{brand2022}, which is not optimized for VCM scenarios.
	When deriving the mask from the YOLO detections (red) as proposed in Sec.~\ref{subsec:VCM optimized mask generation}, the required bitrate can further be reduced compared to the reference case while even increasing the detection accuracy.
	As an oracle test, we also conducted experiments with optimal inference masks derived from the GT data (purple) resulting in a slightly higher detection accuracy at the same bitrate than the VCM-optimized masks.
	This is due to the fact that YOLO does not perfectly find all salient objects in the Cityscapes dataset.
	Thus, those missed objects are transmitted with a worse quality, which ultimately leads to missed detections by the Mask R-CNN that is applied to the coded images.
	From this we can follow that the detection accuracy of the network taken to generate the masks is vital, as missed detections can have severe impact on the whole framework.
	Despite those possible misses, the Mask R-CNN detection accuracy is still higher than for VVC-coded images at all investigated bitrates.
	All in all, utilizing the proposed VCM-optimized mask generation method results in 66.2\,\% bitrate savings over VTM-10.0. With an optimal mask generator, 70.0\,\% of bitrate could be saved.

	\begin{table}[]
		\centering
		\footnotesize
		\caption{Bj\o ntegaard delta values with VTM-10.0 as anchor.}
		{
			\begin{tabular}{lllll|r}
				\toprule           
				\makecell[lb]{\\Marker} & \makecell[lb]{Codec} & \makecell[lb]{Train\\ loss} & \makecell[lb]{Train\\ mask} & \makecell[lb]{Inf. \\ mask} & \makecell[lb]{BDR \\ wAP} \\ \midrule
				\showmark{color5}{Mercedes star}         & VTM-10.0            & -          & - & YOLO & -62.0\,\%\\
				\showmark{color0}{*}         & NCN~\cite{fischer2022_journal}             & \lossTask          & - & - & -41.4\,\%\\
				\showmark{color1}{square*}  & RDONet          & \lossHVS           & var & var & 21.5\,\% \\
				\showmark{color2}{diamond*}  & RDONet          & \lossTask          & var & var & -52.7\,\% \\
				\showmark{color3}{asterisk}  & RDONet          & \lossTask          & var & YOLO & -66.2\,\% \\
				\showmark{color4}{pentagon*}  & RDONet          & \lossTask          & var & GT & -70.0\,\% \\
				\showmarkDashed{color3}{asterisk}  & RDONet          & \lossTask          & GT & YOLO & -77.1\,\% \\
				\showmarkDashed{color4}{pentagon*}  & RDONet          & \lossTask          & GT & GT & -79.5\,\% \\
				\bottomrule
			\end{tabular}
		}
		\label{tab:BD results}
		\vspace{-3mm}
	\end{table}
	
	\def\imageSize{0.50}
	\begin{figure}[!t]%
		\centering
		\begin{tabular}{p{0.45\linewidth}p{0.45\linewidth}}
			\includegraphics[width=\linewidth]{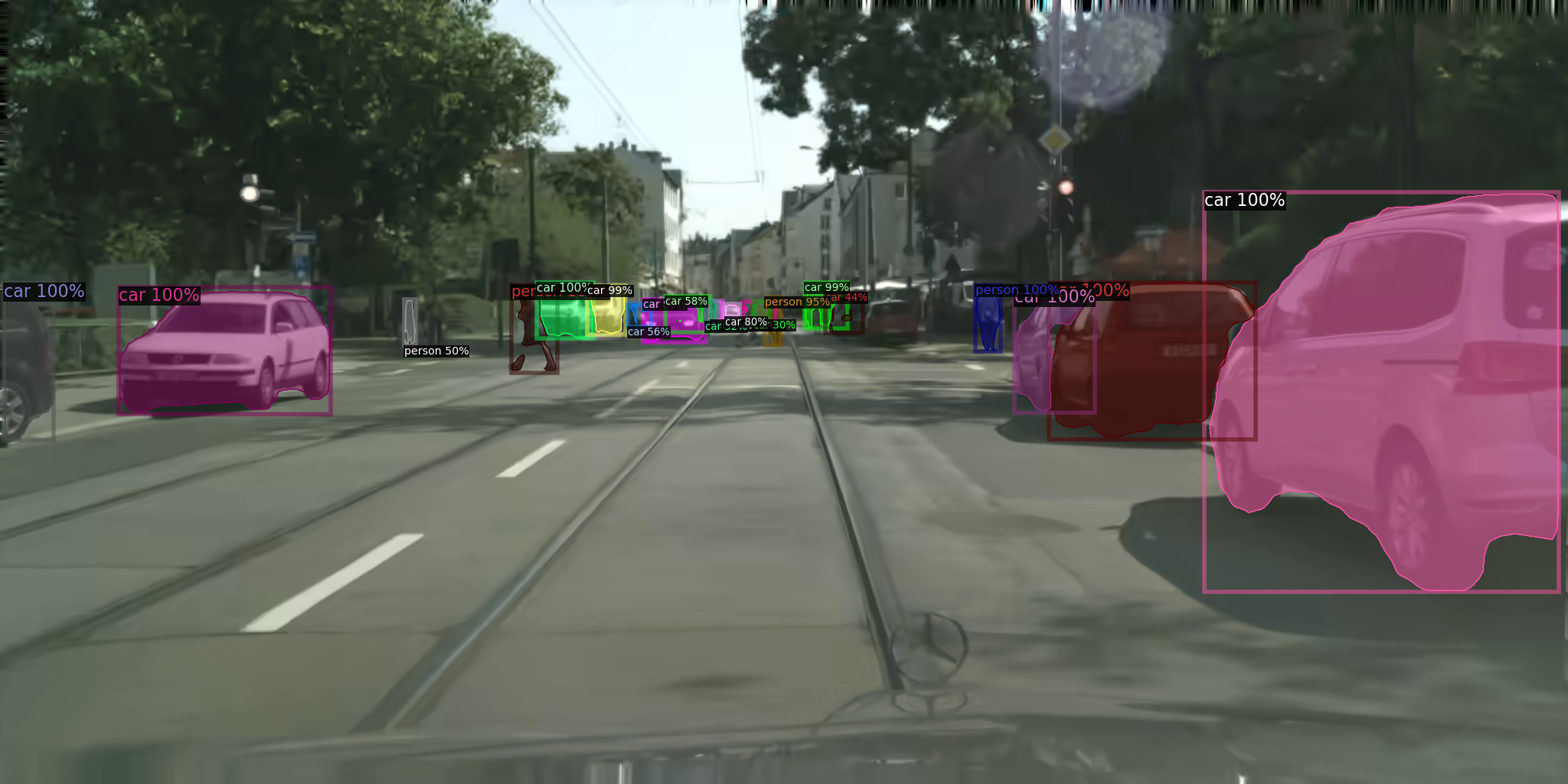} &
			\includegraphics[width=\linewidth]{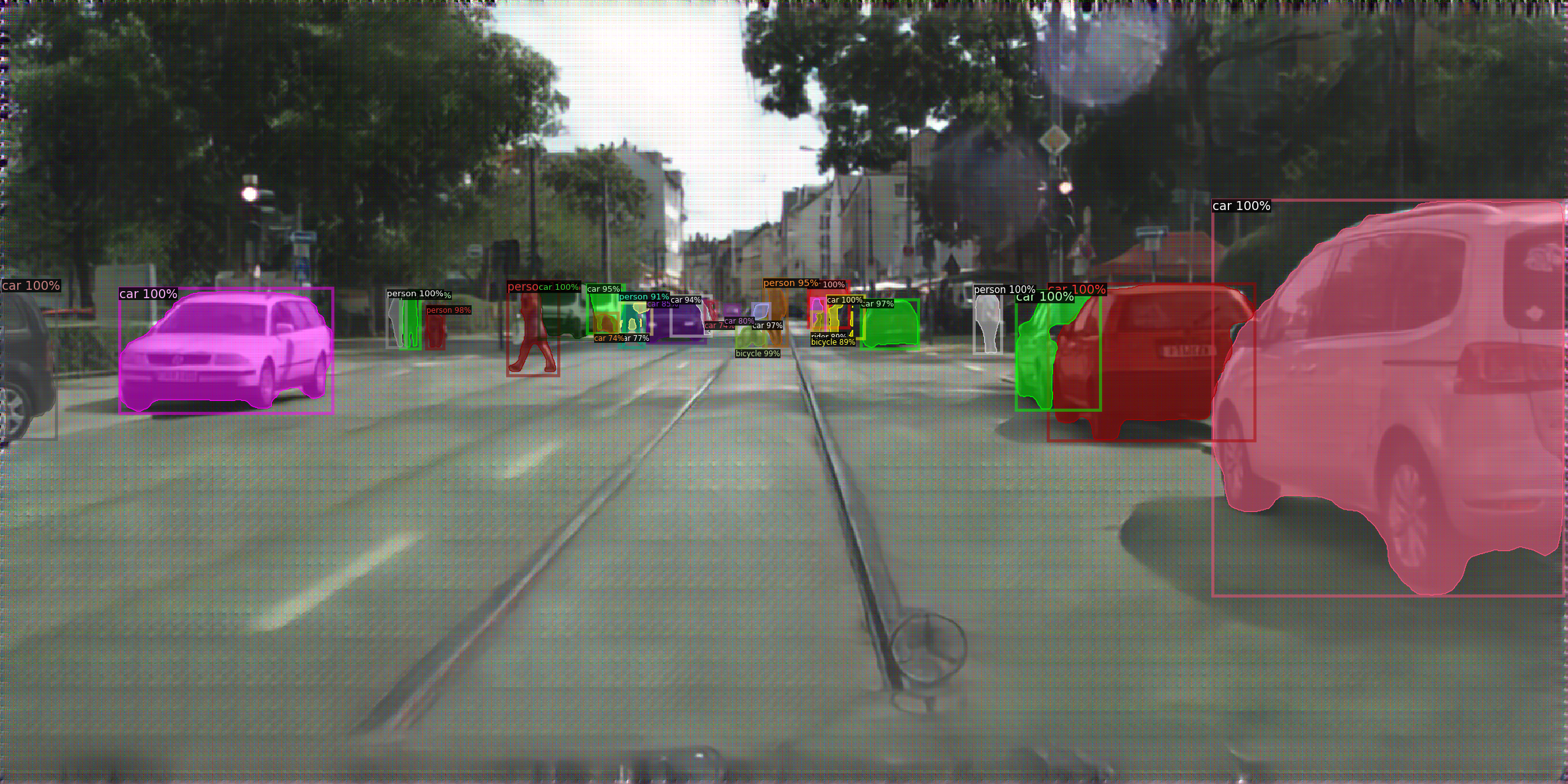} \\
			\footnotesize{\makecell[t]{a) VTM-10.0 (QP=37)\\ @ 0.057 bits per pixel}} &
			\footnotesize{\makecell[t]{b) Train: var mask, Inf.: var mask \\@ 0.078 bits per pixel}} \\
			
			\includegraphics[width=\linewidth]{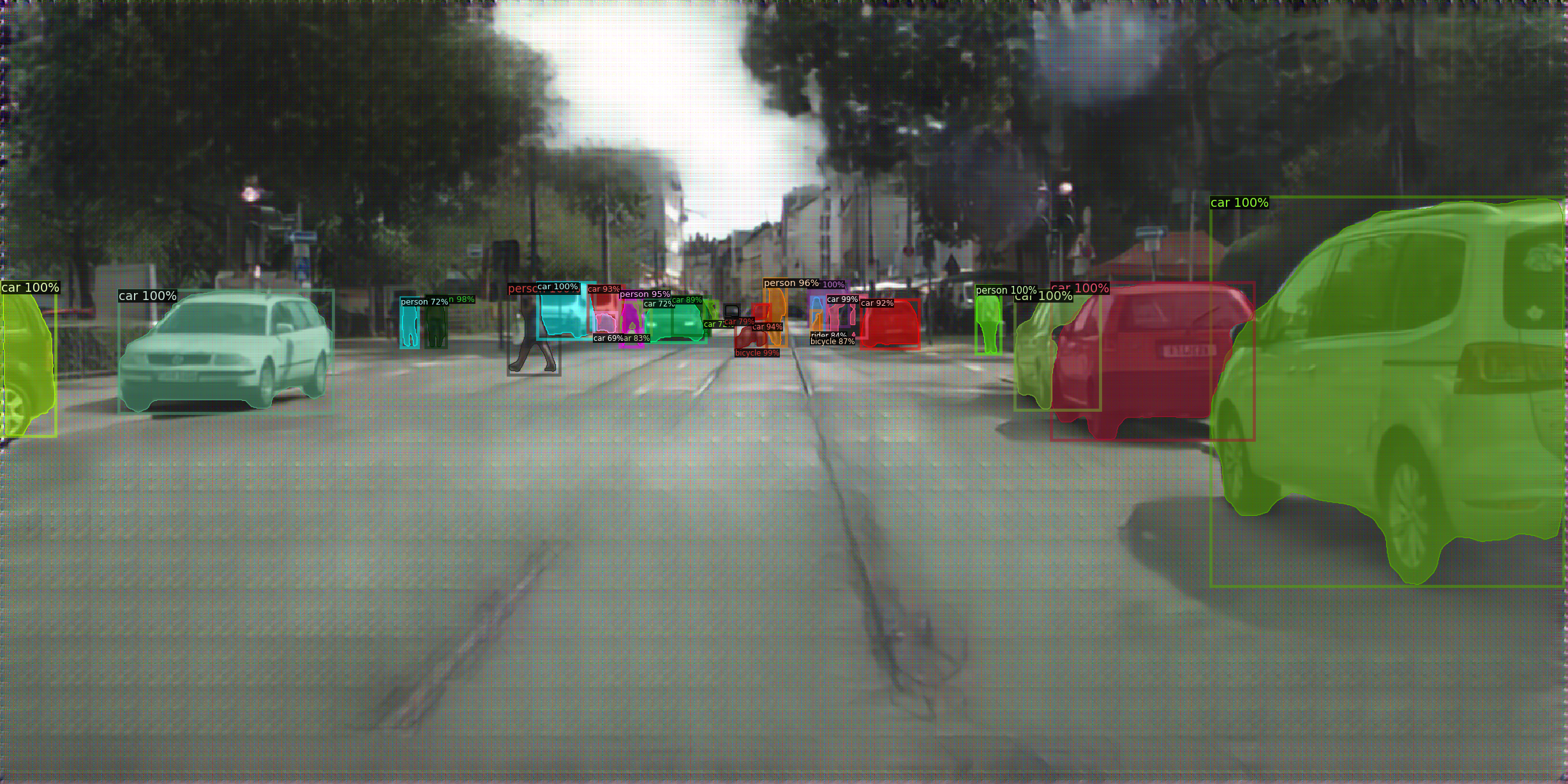} &
			\includegraphics[width=\linewidth]{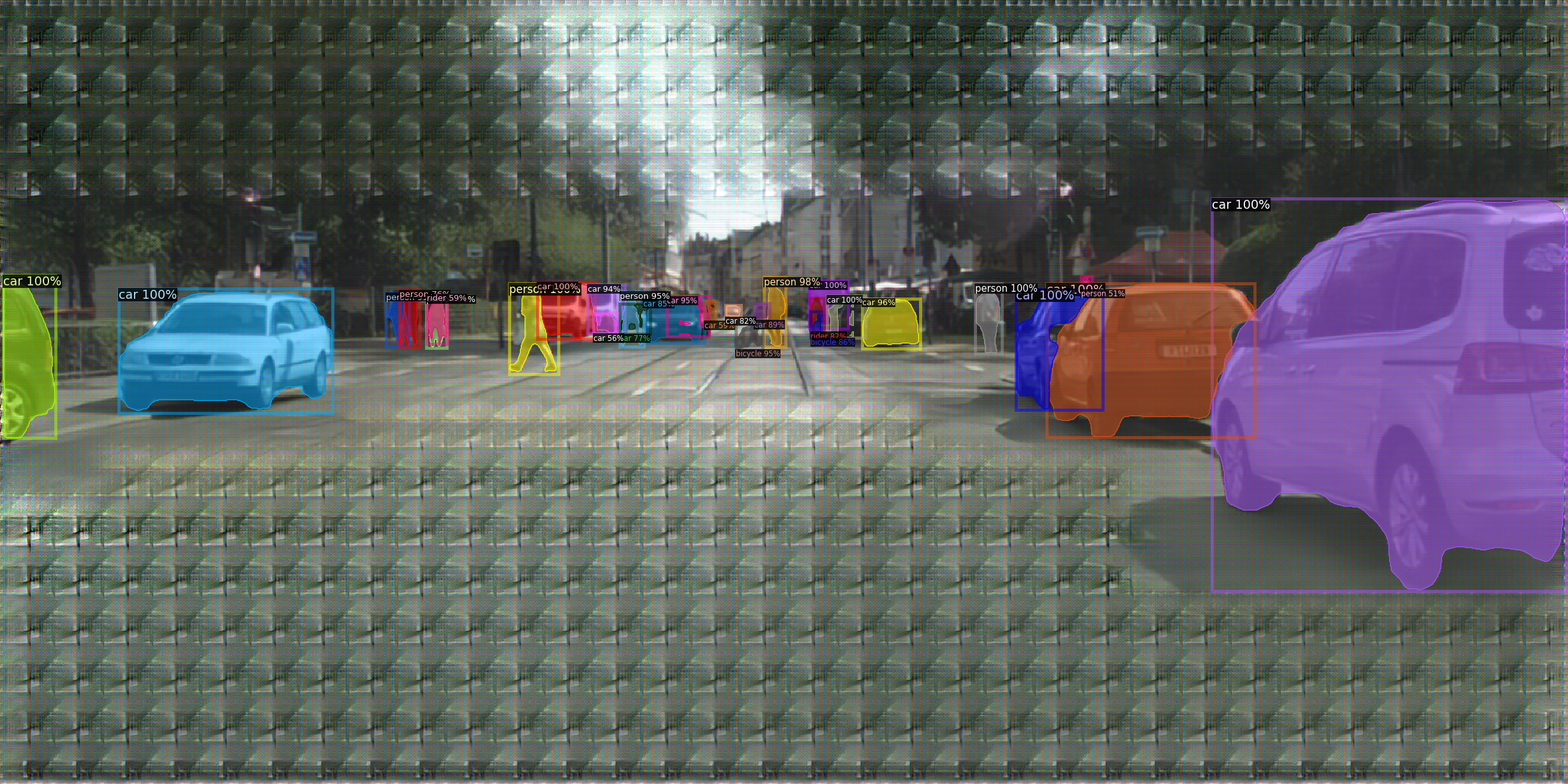} \\
			\footnotesize{\makecell[t]{c) Train: var mask, Inf.: YOLO \\ mask @ 0.059 bits per pixel}} &
			\footnotesize{\makecell[t]{d) Train: GT mask, Inf.: YOLO \\ mask @ 0.047 bits per pixel}} \\
		\end{tabular}
		\vspace{-3mm}
		\caption{Visual results for coding the exemplary Cityscapes image \textit{frankfurt\_000000\_001236\_leftImg8bit} with different RDONet models and the corresponding Mask R-CNN detections. All models were trained with \lossTask\ on the same $\lagrange$ value. Corresponding masks are depicted in Fig.~\ref{fig:exemplary masks}. Best to be viewed enlarged on a screen.}%
		\label{fig:visual results}%
		\vspace{-5mm}
	\end{figure}
	
	\subsection{Influence of Improved Training with GT Masks}
	For the previously shown results, the networks were all trained with the variance-based masks as proposed in~\cite{brand2022}.
	The dashed lines in Fig.~\ref{fig:comparison different mask generation methods} represent the coding behavior when the models are trained with masks derived from the GT data as proposed in Sec.~\ref{subsec: improved training with GT data}.
	The curves show that training the models with the VCM-optimal masks further increases the coding efficiency by reducing the bitrate in non-salient areas.
	In terms of BDR, the model trained on GT-based masks and executed with the YOLO mask during inference achieves bitrate savings of 77.1\,\% over VTM-10.0.
	By that, the proposed framework with RDOnet achieves 15.1 percentage points more BDR savings compared to applying the saliency-driven method proposed in~\cite{fischer2021_ICASSP} with YOLO as saliency detector on VTM-10.0 (brown).
	Our method also clearly outperforms the network proposed in~\cite{le2021_ICASSP} (-33.7\,\% BDR over VTM-8.2), and our previous LSMnet method~\cite{fischer2022_journal}, which adds implicit saliency information derived from the Mask R-CNN features to the coding process (-54.3\,\% BDR over VTM-10.0).

	\subsection{Visual Results}
	
	Fig.~\ref{fig:visual results} gives a visual comparison.
	When coding with the proposed YOLO mask (cf. Fig.~\ref{fig:visual results}c), non-relevant details such as street markers or trees are coded in $\latentRepresentation_3$, and thus with lower quality requiring less rate.
	If the model is also trained with explicit saliency information (cf. Fig.~\ref{fig:visual results}d), the quality is drastically reduced in the non-salient areas.
	The relevant objects in the image are transmitted with high quality and can still be detected by the analysis network.
	
	These visual results also show that the high VCM coding efficiency comes at a price.
	Considering a scenario where, e.g., a human supervisor is supposed to comprehend the detections of the analysis network from the transmitted image.
	This would be possible in the salient areas, but not whether there might be missed objects in areas that have been classified as non-salient during mask generation at the encoder. 
	Therefore, the proposed method can be regarded as an intermediate step between image and feature coding for machines.
	Future research might add a HVS-based regularization term to $\lossTask$, in order to not let the quality in non-salient regions drop below a certain visual quality depending on the practical use case.

	\vspace{-3mm}
	\section{Conclusion}
	
	In this paper we proposed to employ a hierarchical neural image compression network for the VCM context, which transmits the information in multiple latent spaces. 
	To adapt this network to a coding framework with an instance segmentation network as information sink, we end-to-end trained the NCN with the analysis network as discriminator, proposed a VCM-optimized saliency mask generation, and also utilized a mask derived from the GT data to optimally adapt the different latent spaces during training.
	With all our proposed optimizations, our RDOnet model is able to save 77.1\,\% of bitrate over VTM-10.0 at the same detection accuracy.
	Thereby, RDOnet also clearly outperforms existing NCN approaches with one latent space and the reference case when applying the same saliency criterion based on YOLO to VTM-10.0.
	
	\bibliographystyle{IEEEbib}
	\bibliography{/home/fischer/Paper/literature_M2M_communication.bib}
	
\end{document}